\documentclass[prl,superscriptaddress,showpacs,twocolumn]{revtex4}
\usepackage{graphicx}
\begin{document}

\title{Dipole-induced vortex ratchets in superconducting
films with arrays of micromagnets}

\author{C. C. de Souza Silva}
\email{clecio@df.ufpe.br}

\affiliation{INPAC-Institue for Nanoscale Physics and Chemistry, Nanoscale Superconductivity and Magnetism \& Pulsed Fields Group, Katholieke Universiteit Leuven, Celestijnenlaan 200 D, B-3001 Leuven, Belgium}

\affiliation{Departamento de F\'{\i}sica, Universidade Federal de Pernambuco, 50670-901 Recife-PE, Brazil}
\thanks{Permanent address.}

\author{A. V. Silhanek}

\author{J. Van de Vondel}

\author{W. Gillijns}
\affiliation{INPAC-Institue for Nanoscale Physics and Chemistry, Nanoscale Superconductivity and Magnetism \& Pulsed Fields Group, Katholieke Universiteit Leuven, Celestijnenlaan 200 D, B-3001 Leuven, Belgium}

\author{V. Metlushko}
\affiliation{Department of Electrical and Computer Engineering, University of Illinois, Chicago, IL 60607}

\author{B. Ilic}
\affiliation{Cornell Nanofabrication Facility, School of Applied and Engineering Physics, Cornell University, Ithaca, New York 14853}

\author{V. V. Moshchalkov}
\affiliation{INPAC-Institue for Nanoscale Physics and Chemistry, Nanoscale Superconductivity and Magnetism \& Pulsed Fields  Group, Katholieke Universiteit Leuven, Celestijnenlaan 200 D, B-3001
Leuven, Belgium}

\date{\today}

\begin{abstract}

We investigate the transport properties of superconducting films with periodic arrays of in-plane magnetized micromagnets. Two different magnetic textures are studied: a square array of magnetic bars and a close-packed array of triangular microrings. As confirmed by MFM imaging, the magnetic state of both systems can be adjusted to produce arrays of almost point-like magnetic dipoles. By carrying out transport measurements with ac drive, we observed experimentally a recently predicted ratchet effect induced by the interaction between superconducting vortices and the magnetic dipoles. Moreover, we find that these magnetic textures produce vortex-antivortex patterns, which have a crucial role on the transport properties of this hybrid system.

\end{abstract}


\maketitle

Nanoengineered arrays of vortex pinning sites have recently spawned numerous novel phenomena and potential applications based on vortex manipulation in superconductors. These structures are very suitable for tailoring the critical current and equilibrium properties~\cite{Baert95,Raedts04,Lange03} as well as for manipulating the distribution and direction of motion of flux quanta in superconducting devices~\cite{Villegas03,Joris05,Clecio06a,Clecio06b}. Such a control of vortex motion can be achieved when the periodic pinning potential lacks the inversion symmetry in a given direction, in which case any correlated fluctuating force induces a net vortex motion based on the phenomenon known as {\it ratchet effect}~\cite{Reimann_Rev}.

As recently proposed by Carneiro~\cite{Gilson_Ratchet}, a different way to create vortex ratchets can be realized by using in-plane magnetized dots. Here the spatial inversion symmetry is broken not by the shape of the pinning sites but rather by the vortex-magnetic dipole interaction. This dipole-induced ratchet motion depends on the orientation and strength of the local magnetic moments thus allowing one to control the direction of the vortex drift. It is particularly this flexibility to manipulate the vortex motion which makes this kind of pinning potentials attractive for practical applications, although still a clear experimental corroboration is pending. In the present work, we demonstrate in a series of transport experiments that in-plane magnetized dipoles can indeed rectify vortex motion. Moreover, the rectified voltage induced by the ratchet motion depends strongly on temperature and field intensity and is non-zero even at zero field. Our analysis suggests that this behavior results from the interaction between the external field-induced vortices and vortex-antivortex pairs generated by magnetic dipoles.

Our measurements were performed on two samples with different magnetic templates: (i) a 50 nm thick Al film with an on-top square array (period  $a_p=3$ $\mu$m) of Si/Co/Au bars (with thicknesses 5 nm/47 nm/5 nm and lateral dimensions $2.6\times 0.5$ $\mu$m$^2$), labeled Bar-Al, and (ii) a Ge/Pb/Ge trilayer (20 nm/25 nm/5 nm thick) evaporated on top of a close-packed square array of equilateral triangular Co rings (250 nm wide, 23 nm thick, with lateral size 2 $\mu$m and apart 250 nm), labeled Tri-Pb. The insulating layer is deposited between the superconducting and the micromagnet arrays (MMA) to avoid proximity effect. The critical temperature, coherence length, penetration depth, and the first matching field ($H_1$, the field that generates a vortex distribution of one flux quantum per MMA unit cell) were respectively $T_c=1.325$ K, $\xi(0)=87$ nm, $\lambda(0)=130$ nm, and $H_1=0.23$ mT for Bar-Al, and $T_c=7.22$ K, $\xi(0)=33$ nm, $\lambda(0)=35$ nm, and $H_1=0.465$ mT for Tri-Pb.

\begin{figure}[b]
\vspace{-2mm} \centering
\includegraphics[width=8cm]{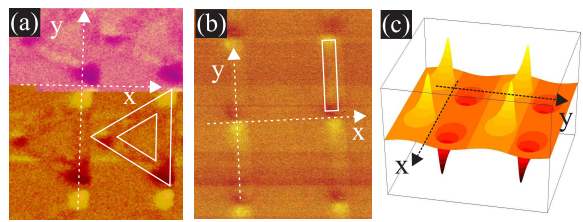}
\caption{\label{MFM} (color online). MFM images of samples Tri-Pb (a) and Bar-Al (b) polarized in the $\hat{y}$ direction. $\hat{x}$ indicates the direction of positive applied current. (c) Calculated pinning potential for the MMA of Bar-Al.}
\end{figure}

Prior to the transport measurements, the MMAs were magnetized with a 500 mT in-plane field oriented along the $\hat{y}$ direction (Fig.~\ref{MFM}).
Owing to their large aspect ratio, the remnant state of all bars (or triangle sides) is single domain~\cite{Seynaeve01}, with the magnetic poles lying very close to the extremities of the bars. This is demonstrated by room-temperature magnetic force microscopy (MFM) measurements (Fig.~\ref{MFM}). As a consequence, the small separation between neighboring bars works as a dipole with an effective magnetic moment pointing against the magnetization of the bars. In sample Tri-Pb a considerably weaker dipole can also be observed at one of the vertices of each triangle. It is worth noticing that the large Curie temperature of Co ($\sim 1400$ K) guarantees that the magnetic states of Fig.~\ref{MFM} remain essentially the same at low temperatures. Moreover, due to a coercive field of $\sim 40$ mT, the small perpendicularly applied magnetic fields ($< 3$ mT), used in our experiments to generate the vortex distributions, are not able to change the magnetic state of the MMAs.

\begin{figure}[t]
\centering
\includegraphics[width=7cm]{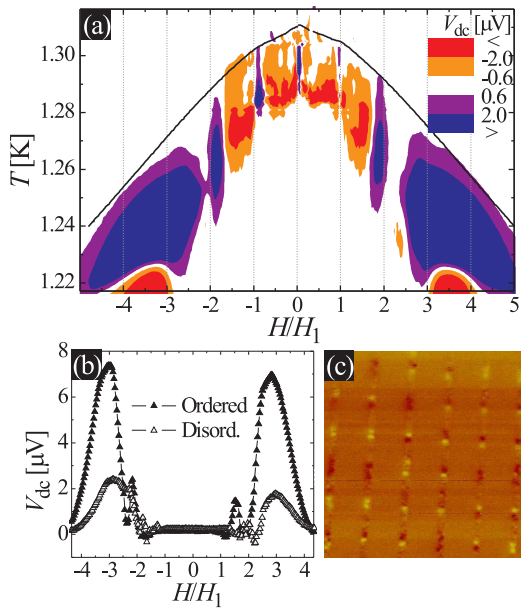}
\caption{\label{HTAl}(color online). (a) Rectified voltage $V_{dc}$ in the temperature-field ($T$-$H$) plane for sample Bar-Al subjected to a 1 kHz sinusoidal ac current of amplitude 0.8 mA. The full line indicate the normal-superconductor transition. (b) $H$ dependence of $V_{dc}$ at $T=1.27$ K in the polarized state (filled dots) and in the disordered state (open dots) depicted in the MFM image (c).}
\end{figure}
Fig.~\ref{MFM}(c) presents the pinning potential $U_{\rm MA}$ generated by the MMA of sample Bar-Al, as felt by a singly-quantized vortex, calculated following standard procedure~\cite{Milosevic03,Gilson05_PRB}. This potential is antisymmetric (along the $\hat{y}$ direction) with respect to the origin of the coordinate system (midpoint between consecutive bars), thus producing the broken inversion symmetry required for the ratchet effect. Since current applied along $+\hat{x}$ ($-\hat{x}$) induces a Lorentz force on a vortex along $-\hat{y}$ ($+\hat{y}$), the asymmetry of this potential may be probed by applying an oscillating current in the $x$ axis (in the case of sample Tri-Pb such a current probes only the dipole-induced asymmetry rather than the broken geometrical symmetry of the triangles). Notice however that $U_{\rm MA}$ changes sign when we are dealing with antivortices rather than vortices, i.e. while vortices drift more easily along $+\hat{y}$, the easy drive for antivortices is $-\hat{y}$, in contrast with the ratchet potentials generated by asymmetric antidots~\cite{Villegas03,Joris05,Clecio06a,Clecio06b}. Because $V_{dc}\propto v_dH$ (with $v_d$ the vortex drift velocity), {\it the measured dc voltage $V_{dc}$ is insensitive to the field polarity}. Thus, a fingerprint of dipole-induced ratchet effect is the presence of mirror-like $V_{dc}(H)$ regions with respect to $H=0$.

\begin{figure}[t]
\centering
\includegraphics[width=7cm]{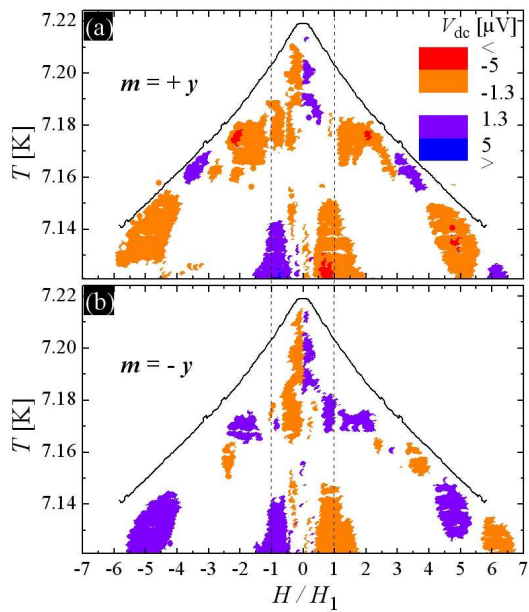}
\caption{\label{HTPb}(color online). Rectified voltage $V_{dc}$ in the $T$-$H$ plane for sample Tri-Pb in the $+\hat{y}$ (a) and
$-\hat{y}$ (b) polarized states for a 1.1 kHz sinusoidal
excitation of amplitude 1 mA. Full lines indicate the normal-superconductor transition.}
\end{figure}

In order to test whether this situation describes our system properly, we apply a sinusoidal ac current while recording the output dc voltage $V_{dc}$ generated by the flow of vortices. For a symmetric pinning landscape like the one provided by a plain film, one should expect a very low $V_{dc}$, as indeed observed in our experiment. In contrast to that, for the polarized states a prominent ratchet drift ($V_{dc}\neq 0$) can be observed as is shown in Figs.~\ref{HTAl} and \ref{HTPb}. The temperature-magnetic field ($T$-$H$) diagrams of sample Bar-Al (Fig.~\ref{HTAl}) clearly demonstrate a symmetrical $V_{dc}(H)$. $T$-$H$ diagrams of sample Tri-Pb with the triangles magnetized along $+\hat{y}$ and $-\hat{y}$ are shown in Figs.~\ref{HTPb}(a) and (b) respectively. In this case mirror-like $V_{dc}(H)$ regions can be also observed but only for $|H|>H_1$. For $|H|<H_1$ and for both magnetic orientations of the dipoles (i.e. it is a common feature in Figs. 3(a) and (b)), $V_{dc}(H)$ is antisymmetric, therefore the signal in this field range is dominated by a fixed (non magnetic) ratchet potential as those generated for instance by small asymmetries of the triangles due to shadow evaporation or a slight misalignment of the bridge axis with respect to the square array.

A major advantage of using magnetic structures to manipulate vortices is the tunability of their magnetic state and, consequently, of their flux pinning properties. This constitutes a valuable tool to control the strength and direction of the ratchet effect. This tuneability becomes apparent when using different magnetic states of the MMAs. In Fig.~\ref{HTPb}, for instance, it is clear that changing the magnetic state of the triangles of sample Tri-Pb from $-\hat{y}$ to $\hat{y}$ induces a sign change in $V_{dc}$. For sample Bar-Al, we have built a disordered state of the bars where most of them are still in the single domain state while others are either completely demagnetized or with opposite magnetization (Fig.~\ref{HTAl}(c)). The $V_{dc}(H)$ curves of both ordered and disordered states exhibit similar features, but the disordered system had a weaker $V_{dc}$ signal as expected (Fig.~\ref{HTAl}(b)). {\it These results represent a clear and unambiguous evidence that the observed ratchet is genuinely magnetic in origin}. The demonstration of vortex ratchets induced by triangular rings is particularly interesting because, whereas the bars can not be demagnetized in a straightforward way, the triangles could be easily induced into a magnetic vortex state~\cite{Alejandro} which erase the total magnetic moment in every element, thus providing a reliable way to switch off the ratchet effect. On the other hand, microbar arrays present a simpler magnetic texture, suitable for understanding the magnetic-dipole ratchet mechanism.

Let us now analyze some features observed in sample Bar-Al in more detail. In the $T$-$H$ diagram of Fig.~\ref{HTAl}, one can clearly see at lower matching fields and even at $H=0$ narrow regions of positive voltage embedded in larger regions of negative voltage. Such a zero-field ratchet effect suggests the existence of vortex-antivortex (V-AV) pairs generated by the MMA which interact with the underlying asymmetric potential giving rise to the observed ratchet signal with no applied field.

Under these circumstances, a V-AV excitation with vorticity $(L_v,L_{av})$ introduces a change in the superfluid kinetic energy, given by $E_{\rm k}(R) = L_vL_{av}(V(R)-V(0)) + \frac{1}{2}(L_v-L_{av}) V(0)$ (where $V(R)$ is the attractive interaction energy of a singly quantized V-AV pair and $R=|\vec{r}_v-\vec{r}_{av}|$), as well as a change in the condensate energy $E_{\rm c} = (L_v+L_{av})\pi r_0^2d{H_c^2}/{8\pi}$ (with $r_0$ the effective vortex core radius and $H_c$ the thermodynamical critical field)~\cite{MinhagenRMP}. By symmetry considerations one can show that the V-AV pair is aligned with the $y$ axis and $\vec{r}_{av}=-\vec{r}_{v}$ (and thus $R=2y_{v}$). The total change in the Gibbs free energy caused by the appearance of a pair can thus be expressed in terms of $y_{v}$ only:
\begin{eqnarray}
\Delta{G}_{L_v,L_{av}}(y_{v})&=&(L_v+L_{av})U_{\rm MA}(y_{v}) +
E_{{\rm k}}(2y_v)\nonumber \\
&+& E_{{\rm c}}-(L_v-L_{av})\Phi_0Hd,
\end{eqnarray}
where the last term accounts for the interaction with the external field. The index $L_v,L_{av}$ in $E_k$ and $E_c$ is omitted for clarity. $V(R)$ and $r_0$ were calculated within Clem's variational approach~\cite{Clem74} for the Ginzburg-Landau equations and assuming $R\ll\lambda^2/d$ \cite{ClecioUnpub}. Let us assume that at $H=0$ each magnet stabilize a $(L_0,L_0)$ pair, thus generating a lattice of neutral V-AV pairs. Accordingly, $E_{{\rm k}}(R)$ must account for the interaction of a vortex with the whole array of antivortices and vice-versa. The zero-field vorticity $L_0$ and the equilibrium pair position $y_0$ can be estimated by minimizing $\Delta{G}_{L_v,L_{av}} (y_v)$ at $H=0$. The solution is thermodynamically stable if $\Delta{G}_{L_0,L_{0}}(y_{0})<0$. For the parameters of sample Bar-Al, we find $L_0=6$ at $T=0.9T_c$. Although this is a rough estimate, it is a good indication that the magnetic template of sample Bar-Al indeed generate V-AV pairs~\cite{Comment}.

\begin{figure}[b]
\centering
\includegraphics[width=7.5cm]{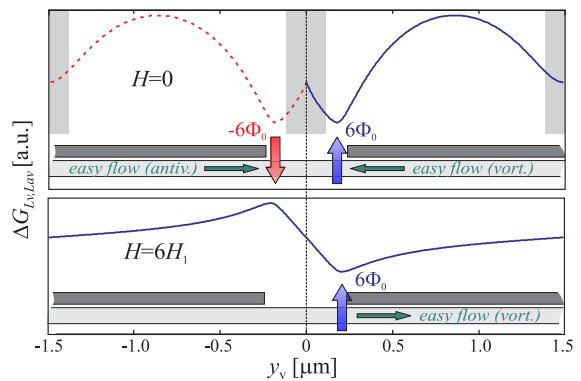}
\caption{\label{DeltaG}(color online). Total energy change of stable V-AV pair configurations as a function of pair position $(y_v,y_{av}=-y_v)$ at $H=0$ and $6H_1$, calculated using for the parameters of sample Bar-Al at $T=0.9T_c$. The curves are shown from the point of view of vortices (full lines) and antivortices (dashed lines), in one unit cell of the MMA. The cartoons illustrate the equilibrium V (upward arrows)-AV (downward arrows) configurations. Small arrows indicate the direction of easy motion of vortices and antivortices. Shaded areas correspond to pair separation $R<\xi(T)$, where, roughly, V-AV annihilation takes place.}
\end{figure}

We can now go a step further in this analysis and determine the zero-field ratchet motion. The upper panel of Fig.~\ref{DeltaG}
shows $\Delta{G}_{L_v,L_{av}}(y_{v})$ curve at $H=0$ in a unit cell of the MMA. Shaded areas correspond to V-AV annihilation zones, where our model is no longer valid. A positive current, i.e. along $+\hat{x}$ (out-of-the-page) direction, force both vortex and antivortex to approach each other while a negative current tear them away from each other. Notice that the symmetry constraint $y_{av}=-y_v$ is implicit here. Interestingly the resulting energy profile exhibits a steeper slope for motion of vortices to the right ($+\hat{y}$) and antivortices to the left ($-\hat{y}$), and therefore expanding the V-AV pair is harder than compressing it. As a consequence, an oscillating current induces more vortex activity during the positive half-cycles, thus resulting in a positive dc voltage, as that observed in the $T$-$H$ diagram of sample Bar-Al at $H=0$. However, at a high enough field $(H\geq L_0H_1)$, all antivortices are annihilated. In this case, the V-AV attraction term is absent and the vortex energy profile (Fig.~\ref{DeltaG} lower panel) is determined by the bare vortex-magnetic dipole potential, which has gentler slope for vortex motion along $+\hat{y}$. Thus, in such high fields, an oscillating current induces $V_{dc}<0$. 

The ratchet motion at $H=0$ relies on a very symmetric V-AV configuration and, in general, it can not be expected to hold for small incommensurate fields. Indeed, detuning the field from $H=0$ lead to a negative voltage sign as observed in Fig. 2. A possible explanation is that at small incommensurate fields the extra vortices or antivortices are loosely attached to the pinning potential and thus are able to move at lower critical forces. In this sense, the extra vortices dominate the dynamics in such a way that the ratchet signal is determined by their mobility in one direction or the other. For instance, at small positive fields, an additional vortex will be attracted by a trapped antivortex on its right side while being repelled by a trapped vortex on its left thus favoring motion in the $+\hat{y}$ direction. This leads to a negative $V_{dc}$, in agreement with our experimental observations presented in Fig.~\ref{HTAl} (a). However, the detailed dynamics of vortices and antivortices in an array of in-plane magnetized dipoles is still an open problem. For instance, while multiple sign inversions in non-magnetic ratchets is well explained by vortex-vortex repulsion ~\cite{Clecio06b}, here V-AV attraction is a new ingredient that can play an important role in the sign inversions of Fig. 2 (a). The extreme case occurs at $H=0$, where V-AV attraction is the ultimate factor determining the ratchet drift as demonstrated in Fig. 4. Clearly, further theoretical analysis using, e.g., the time dependent Ginzburg-Landau equations would be of great importance to elucidate the details of the vortex dynamics in these systems.

In summary, we provided the experimental evidence that a periodic array of in-plane magnetized dipoles induces a new kind of ratchet motion generating a dc voltage that is independent of field polarity. We showed that the magnetic elements of our samples are able to generate vortex-antivortex pairs. The presence of these pairs causes a dramatic change in the magnetic ratchet effect not foreseen in earlier work~\cite{Gilson_Ratchet}. For instance, at zero magnetic field the dipole-induced vortex-antivortex pairs generate a positive voltage readout in contrast with the expected negative voltage sign for a single vortex interacting with the dipoles. Such vortex-antivortex mixtures in hybrid superconductor-ferromagnetic systems presented here is an interesting example of the ratchet effect of binary mixtures, where the drift direction is assisted by the interaction between particles of different species~\cite{Savelev}. Other examples can be found in different physical systems such as ion mixtures in cell membranes~\cite{Cabral} and Abrikosov-Josephson vortex mixtures in cuprate high-$T_c$ superconductors~\cite{Cole}. 

This work was supported by the K.U.Leuven Research Fund GOA/2004/02 and FWO programs. C.C.S.S. acknowledges the support of the Brazilian Agencies FACEPE and CNPq. A.V.S. is grateful for the support from the FWO-Vlaanderen.


\begin{thebibliography}{1}
\bibitem{Comment} As a test, we analyzed the hybrid Pb/Co sample studied in Ref.~\onlinecite{VanBael01} and found $L_0=1$, in agreement with their direct imaging results. Using a similar procedure, Milosevic and Peeters~\cite{Milosevic04} have also found $L_0=1$ for this particular experiment.
\end{thebibliography}
\end{document}